\documentclass[usenatbib]{mnras}
\usepackage{graphicx}
\usepackage{dcolumn}
\usepackage{amssymb} 
\usepackage{amsmath}
\usepackage{ctable}
\usepackage{float}
\usepackage{bm}
\usepackage{float}
\usepackage{epsfig}
\usepackage{epstopdf}
\usepackage{epsf,color}
\usepackage{comment}
\usepackage{aas_macros}
\usepackage{url}
\usepackage{soul}
\usepackage{widetext}
\usepackage{color, colortbl}
\newcommand{\cmnt}[1]{}

\usepackage[normalem]{ulem}
\usepackage{color}

\usepackage{tgtermes}
\newcommand{\cii}{\ion{C}{ii}}

\newcommand{\oiii}{\ion{O}{iii}}
\newcommand{\oii}{\ion{O}{ii}}

\newcommand{\sii}{\ion{S}{ii}}

\title[$\oiii$ 52 $\mu$m forecast]{The prospects for observing [\oiii] 52 micron emission from galaxies during the Epoch of Reionization}

\author[S. Yang et al.]{
Shengqi Yang,$^{1}$\thanks{E-mail:sy1823@nyu.edu}
Adam Lidz,$^{2}$
Gerg\"o Popping$^{3}$
\\
$^{1}$Center for Cosmology and Particle Physics, Department of Physics, New York University, 726 Broadway, New York, NY, 10003, U.S.A.\\
$^{2}$Department of Physics and Astronomy, University of Pennsylvania, 209 South 33rd Street, Philadelphia, PA 19104, USA\\
$^{3}$European Southern Observatory, Karl-Schwarzschild-Strasse 2, D-85748, Garching, Germany
}

\date{Accepted XXX. Received YYY; in original form ZZZ}

\pubyear{2021}

\begin{document}
\label{firstpage}
\pagerange{\pageref{firstpage}--\pageref{lastpage}}
\maketitle

\begin{abstract}
The [\oiii] 88 $\mu$m fine structure emission line has been detected into the Epoch of Reionization (EoR) from star-forming galaxies at redshifts $6 < z \lesssim 9$ with ALMA. These measurements provide valuable information regarding the properties of the interstellar medium (ISM) in the highest redshift galaxies discovered thus far. The [\oiii] 88 $\mu$m line observations leave, however, a degeneracy between the gas density and metallicity in these systems. Here we quantify the prospects for breaking this degeneracy using future ALMA observations of the [\oiii] 52 $\mu$m line. Among the current set of ten [\oiii] 88 $\mu$m emitters at $6 < z \lesssim 9$, we forecast 52 $\mu$m detections (at 6-$\sigma$) in SXDF-NB1006-2, B14-6566, J0217-0208, and J1211-0118 within on-source observing times of 2-10 hours, provided their gas densities are larger than about $n_{\mathrm{H}} \gtrsim 10^2-10^3$ cm$^{-3}$. 
Other targets generally require much longer integration times for a 6-$\sigma$ detection. Either successful detections of the 52 $\mu$m line, or reliable upper limits, will lead to significantly tighter constraints on ISM parameters. The forecasted improvements are as large as $\sim 3$ dex in gas density and $\sim 1$ dex in metallicity for some regions of parameter space. We suggest SXDF-NB1006-2 as a promising first target for 52 $\mu$m line measurements. We discuss how such measurements will help in understanding the mass metallicity relationship during the EoR.
\end{abstract}




\section{Introduction}

Recent ALMA observations of atomic fine structure emission lines have provided spectroscopic redshifts for galaxies into the EoR at $6 < z \lesssim 9$ and started to probe their ISM properties (e.g., \citealt{2015ApJ...807..180W,2016Sci...352.1559I,2018Natur.553...51M}). Specifically, these measurements constrain the internal structure, dynamics, ionization state, and gas phase metallicity in some of the first galaxies. The fine structure line observations can be further combined with rest-frame ultraviolet (UV) estimates of the star-formation rates (SFRs) in these galaxies and infrared (IR) determinations of stellar mass, allowing one to study correlations between gas content and stellar populations into the EoR. These, in turn, give crucial empirical guidance for models of galaxy formation and help to determine the properties of the sources that reionized the universe. 

More specifically, this paper focuses on [\oiii] fine-structure emission lines from the EoR, which probe the gas phase metallicity and density in the HII regions in these galaxies. First, these observations help in understanding the chemical enrichment history in early galaxy populations. Furthermore, in lower redshift galaxy samples, there is a well-established correlation between gas phase metallicity and stellar mass \citep{1979A&A....80..155L,2004ApJ...613..898T}: this is thought to reflect the impact of outflows which drive gas and metals out of the shallow potential wells of small mass galaxies but have less affect in larger galaxies. The recent [\oiii] measurements, combined with stellar mass estimates, start to study whether these correlations hold and/or evolve into the 
EoR (\cite{2020ApJ...903..150J}, hereafter Jones2020). Next, the gas density measurements are relevant for understanding the internal structure and escape fraction of ionizing photons from these galaxies (e.g. \citealt{2013ApJ...770...76B,2014ApJ...788..121K,2015MNRAS.453..960M,2019MNRAS.486.2215K}). The escape fraction plays a critical, yet highly uncertain, role during cosmic reionization.

Thus far, ALMA has detected the [\cii] 158 micron emission line and the [\oiii] 88 micron line from tens of $6 < z \lesssim 9$ galaxies (e.g., \citealt{2016ApJ...829L..11P,2017ApJ...837L..21L,2017A&A...605A..42C,2018Natur.553..178S,2018MNRAS.478.1170C,2018ApJ...854L...7C,2018Natur.557..392H,2019PASJ...71...71H,2019ApJ...874...27T,2020ApJ...896...93H,2019ApJ...881...63N}). Intriguingly, in some cases the [\oiii] luminosities from this sample exceed those of local galaxies \citep{2014A&A...568A..62D} with the same SFRs \citep{2018MNRAS.481L..84M}.
The ratio between the [\oiii] 88 $\mu$m and [\cii] luminosity is also larger than in local galaxies \citep{2020ApJ...896...93H}. In short, the [\oiii] 88 $\mu$m line is a bright and promising tracer of reionization-era galaxies. 

Motivated by the ALMA [\oiii] 88 $\mu$m detections and their future promise, we developed a first-principles analytic model for [\oiii] emission in \cite{2020MNRAS.499.3417Y}, (hereafter Yang2020). We leave the more complex modeling required for studying [\cii] emission (e.g. \citealt{2019MNRAS.489....1F,2019MNRAS.487.5902K}) to future work.
The Yang2020 model determines the [\oiii] luminosity from galaxies with a given SFR, metallicity, gas density, and ionizing spectral shape. Briefly, in these calculations we first compute the total volume in HII regions across each galaxy and the [\oiii] fraction within these regions. 
We then determine the level populations in the different fine-structure states and the resulting line luminosities. We cross-checked these calculations against \textsc{CLOUDY} \citep{2017RMxAA..53..385F} simulations and find that they agree to better than 15\% accuracy across a broad range of model parameters. We then applied the model to derive bounds on the gas phase metallicity and density in the HII regions from the current ALMA sample of 88 $\mu$m detections and measurements of their luminosity to SFR. (This is denoted herein as $L_{10}$/SFR since the 88 $\mu$m transition is between the first excited level and the ground state, i.e. it is a $1 \rightarrow 0$ transition). 

An important degeneracy is left, however, between the metallicity and gas density from the 88 $\mu$m and SFR measurements alone (Yang2020). At high densities, $n_\mathrm{H} \gtrsim 10^2-10^3$ cm$^{-3}$, collisional de-excitations become important and it is impossible to distinguish galaxies with high density and metallicity from those with lower density and metallicity, since the line luminosity drops with increasing density and/or decreasing metallictiy. 
The 88$\mu$m and SFR measurements alone yield only an upper bound on gas density, $n_\mathrm{H}$, and a lower bound on metallicity, $Z$.

Previous work suggests that future ALMA measurements of the [\oiii] 52 $\mu$m transition may help to break this degeneracy (Jones2020,Yang2020). This line arises from transitions between the second excited and first excited fine structure levels in [\oiii] (and so the luminosity in this line is denoted hereafter as $L_{21}$). As noted in these previous studies, the ratio between the 52 $\mu$m and 88 $\mu$m emission ($L_{21}/L_{10}$) provides a powerful {\it density diagnostic} (e.g. \citealt{2011piim.book.....D}), since the lines have different critical densities and their ratio hence depends on the importance of collisional de-excitations. As the energy splitting between these fine-structure states is small compared to the temperature of the HII region gas, the line ratio is insensitive to the temperature of the gas. Further, the lines arise from the same ion and so the ratio does not depend on the ionization state of the gas, nor appreciably on its metal content.

Jones2020 also considered the prospects for detecting 52 $\mu$m emission with ALMA from some of the current sample of [\oiii] 88 $\mu$m emitting galaxies at $6 < z \lesssim 9$, finding that some of these sources are detectable in reasonable observing times. Here we extend the work in these previous studies to forecast the quantitative improvements on the gas density and metallicity error bars that will be enabled by new 52 $\mu$m measurements in the future. Furthermore, we consider the implications of these improvements for our understanding of the mass-metallicity relationship during the EoR, building off of the earlier work on using the [\oiii] 88 $\mu$m line to constrain this important correlation in Jones2020 and our previous analytic model in Yang2020.

The plan of this paper is as follows. In \S \ref{sec:2} we describe the current ten ALMA [\oiii] targets at $6<z\lesssim9$. \S \ref{sec:3} reviews the [\oiii] 88 $\mu$m and 52 $\mu$m analytic model introduced in Yang2020. In \S \ref{sec:OT} we consider the $n_\mathrm{H}-Z$ parameter space allowed by current $L_{10}/$SFR observations and compute the range of observing times required to achieve $6-\sigma$ 52 $\mu$m detections using the ALMA sensitivity calculator. In \S \ref{sec:5} we perform Monte Carlo Markov Chain (MCMC) calculations to quantify the parameter space improvements expected towards
four promising 52 $\mu$m targets. We show that both successful detections and upper limits on the 52 $\mu$m signal will significantly tighten the gas density posteriors and also bring stronger constraints on the ISM metallicities. We compare the ISM metallicity constraints from Jones2020, Yang2020, and the future improvements enabled by joint [\oiii] 88 $\mu$m and 52 $\mu$m fits in \S \ref{sec:MZ}. We discuss how these observations can advance our understanding of the mass-metallicity relationship during the EoR. We conclude in \S \ref{sec:Conclusion}. 

\section{Data}\label{sec:2}

We consider the sample of nine ALMA [\oiii] 88 $\mu$m detections plus one upper limit at $6 <z \lesssim 9$, published in the current literature, and summarized in Table~\ref{tb:data}. These include one gravitationally-lensed galaxy at $z=9.1$ \citep{2018Natur.557..392H} (MACS1149-JD1), a lensed Y-band drop-out galaxy A2744\_YD4 at $z \sim 8$ \citep{2017ApJ...837L..21L}, and the $z \sim 8$ Y-dropout Lyman break galaxy (LBG), MACS0416\_Y1 \citep{2019ApJ...874...27T}. At $z\sim7$, [\oiii] from one Ly$\alpha$ emitter is detected in a follow up measurement carried out by \cite{2016Sci...352.1559I}. In addition, the LBG B14-65666 is measured by \cite{2019PASJ...71...71H}, and the star forming galaxy BDF-3299 is detected in \cite{2017A&A...605A..42C}, each near $z \sim 7$. Three luminous LBGs, J1211-0118, J0235-0532, and J0217-0208 at $z\sim6$ are presented by \cite{harikane2020large}. The galaxy SFRs summarized in Table~\ref{tb:data} can also be found in \cite{harikane2020large}. We use the Salpeter IMF based results from this study, rather than the ones in \cite{2020ApJ...896...93H} which assume a Charbier IMF, for consistency with our model. Finally, a non-detection of [\oiii] from the Ly$\alpha$ emitting galaxy z7\_GSD\_3811 at $z=7.7$ is reported in \cite{2020arXiv201113319B}.

\begin{table*}
\centering
\begin{tabular}{lccccccccc}
\hline
&&&& \\[-1em]
Name&Redshift&$\frac{\mathrm{SFR}}{[M_\odot/\mathrm{yr}]}$&$\log\frac{M_*}{[M_\odot]}$&$\frac{L_\mathrm{10}}{[L_\odot]}$&Band&$\frac{L_{21}}{L_{10}}$&$\frac{\sigma_{L_{21}}}{[L_\odot]}$&Observing Time/[hour]&ref\\
&&&& \\[-1em]
(1)&(2)&(3)&(4)&(5)&(6)&(7)&(8)&(9)&(10)\\
&&&& \\[-1em]
\hline
&&&&\\[-1em]
MACS1149-JD1&9.110&$4.2^{+0.8}_{-1.1}$&$9.03^{+0.17}_{-0.08}$&$(7.4\pm1.6)$E7&-&2.33,0.56&-&Not Observable&H18    \\
&&&&\\[-1em]
A2744-YD4&8.382&$20.4^{+17.6}_{-9.5}$&$9.29^{+0.24}_{-0.18}$&$(7.0\pm1.7)$E7&9&5.79,0.58&5.2E8&5.9E2,2.0E3,5.9E4&L17   \\
&&&&\\[-1em]
MACS0416-Y1&8.312&$57.0^{+175.0}_{-0.2}$&$8.38^{+0.11}_{-0.02}$&$(1.2\pm0.3)$E9&9&5.23,0.55&3.2E10&9.4E3,3.1E4,8.5E5&T19\\
&&&&\\[-1em]
SXDF-NB1006-2&7.215&$193^{+155}_{-92}$&$8.54^{+0.79}_{-0.22}$&$(9.9\pm2.1)$E8&9&4.75,0.57&3.2E8&1.73,5.51,1.2E2&I16 \\
&&&&\\[-1em]
B14-65666&7.168&$200^{+82}_{-38}$&$8.89^{+0.05}_{-0.04}$&$(3.4\pm0.4)$E9&9&2.56,0.56&8.4E8&3.35,9.01,69.95&H19 \\
&&&&\\[-1em]
BDF-3299&7.109&5.7&$9.3\pm0.3$&$(1.8\pm0.2)$E8&9&1.28,0.55&3.6E8&8.7E2,1.7E3,4.7E3&C17   \\
&&&&\\[-1em]
J0217-0208&6.204&153&$10.36\pm0.36$&$(8.5\pm2.0)$E9&10&0.86,0.54&1.2E9&9.26,13.98,23.49&H20  \\
&&&&\\[-1em]
J0235-0532&6.090&86&$10.36\pm0.36$&$(3.8\pm0.3)$E9&10&0.90,0.55&1.4E9&60.06,92.56,1.6E2&H20  \\
&&&&\\[-1em]
J1211-0118&6.029&136&$10.61\pm0.36$&$(4.8\pm0.7)$E9&10&1.32,0.55&8.7E8&6.84,13.64,39.42&H20 \\
&&&&\\[-1em]
z7\_GSD\_3811&7.664&$>$8.4&$>8.20$&$<$1.6E8&10&7.16,0.56&$<$2.2E8&$>$13.90,$>$47.83,$>$2.3E3&B20\\
\hline
\end{tabular}
\caption{Summary of the ALMA high redshift [\oiii] galaxies studied in this work and the prospects for future 52 $\mu$m detections. 
The columns in the table give: (1) Object names. (2) Redshifts determined from Lyman-$\alpha$, the Lyman break, rest-frame UV absorption
lines, [\cii] 158 $\mu$m, or [\oiii] 88 $\mu$m. (3) The total SFRs inferred from UV and IR luminosities. (4) Stellar mass inferred from SED fits or UV magnitudes. The stellar mass of BDF-3299 is determined in \protect\cite{2015MNRAS.451L..70M}. The stellar masses of J0217-0208, J0235-0532, and J1211-0118 are estimated through an empirical stellar mass versus UV magnitude relationship for $z\sim6$ Lyman break galaxies (LBGs) \protect\citep{2016ApJ...825....5S}. Other stellar mass references are provided in column (10). (5) [\oiii] 88 $\mu$m luminosities. (6) The ALMA frequency bands capturing the redshifted 52 $\mu$m emission line from each galaxy. (7) The range of [\oiii] 52 $\mu$m to 88 $\mu$m luminosity ratios allowed by the $L_{10}/\mathrm{SFR}$ measurements.
(8) The 1-$\sigma$ noise on the 52 $\mu$m (velocity-integrated) luminosity for a 10-hour on-source ALMA measurement. (9) The on-source observing time 
required to achieve a SNR$=6$ [\oiii] 52 $\mu$m detection. The first/second/third entry correspond to the observing time under the maximum/average/minimum $L_{21}/L_{10}$ cases. (10) References for the [\oiii] 88 $\mu$m measurements, the SFR, and stellar masses. H18: \protect\cite{2018Natur.557..392H}, L17: \protect\cite{2017ApJ...837L..21L}, T19: \protect\cite{2019ApJ...874...27T}, I16: \protect\cite{2016Sci...352.1559I}, H19: \protect\cite{2019PASJ...71...71H}, C17: \protect\cite{2017A&A...605A..42C}, H20: \protect\cite{2020ApJ...896...93H}, B20: \protect\cite{2020arXiv201113319B}.}\label{tb:data}
\end{table*}

\section{Model}\label{sec:3}

Our aim is to forecast the expected SNR for 52 $\mu$m emission line observations of this sample of galaxies as well as the improvements expected for the ISM parameter constraints. In order to do this we need to account for current uncertainties in the [\oiii] 88 $\mu$m luminosities and SFRs, and we also need to span the allowed ISM parameter space.   

To accomplish this, we turn to the Yang2020 [\oiii] emission model. In brief, this model treats the ionizing output of each galaxy as concentrated into a single effective source of ionizing radiation at the center of a spherically symmetric HII region, which is in photo-ionization equilibrium. The rate of hydrogen ionizing photons emitted by this source is given by $Q_\mathrm{HI}$ and is determined by the galaxy's SFR, stellar metallicity, and IMF. Although in reality the [\oiii] emission arises from a complex ensemble of discrete HII regions distributed across the galaxy, our simplified treatment -- with a single effective HII region -- should provide an accurate prediction of the total [\oiii] luminosity summed over all of the HII regions in the galaxy. We adopt a \textsc{Starburst99} population synthesis stellar spectrum \citep{1999ApJS..123....3L} with a continuous SFR, a Salpeter IMF \citep{1955ApJ...121..161S}, and an age of 10 Myr throughout, in which case the doubly-ionized oxygen fraction is close to unity throughout the HII region for the SFRs considered here. As discussed in Yang2020, we don't expect the precise choice of stellar spectrum here to significantly impact our results. For simplicity in making our forecasts we ignore variations in the gas density and metallicity across each galaxy (see Yang2020 for extensions to this and further discussion). In this case the gas density is characterized by a single number, $n_\mathrm{H}$, across each galaxy, while the metallicity is described by the parameter $Z$.

For simplicity, our baseline model assumption is that the stellar metallicity matches the gas phase metallicity. Note, however, that \cite{2016ApJ...826..159S} find evidence for super-solar oxygen to iron abundance ratios in $z \sim 2-3$ LBGs and argue that this is a natural consequence of chemical enrichment dominated by core-collapse supernovae, as would also be expected at the $z \geq 6$ redshifts in our sample. Since iron largely controls the stellar opacity and mass loss, this case can be roughly described by adopting a lower stellar metallicty for a given gas phase metallicity. 
To account for this, we therefore consider also an alternate case in which the stellar metallicity is a factor of 5 smaller than the gas-phase metallicity. The reduced stellar metallicity in this scenario increases the ionizing photon output for a given SFR, and thus enhances the [\oiii] emitting volume, relaxing the gas-phase metallicity bound from the $L_{10}/{\rm SFR}$ measurements (see Yang2020 and further discussion below).

The Yang2020 model then solves for the fine structure level populations in the three-level atom approximation, accounting for radiative de-excitations, collisional excitations and de-excitations, and sub-dominant radiative trapping effects (computed in the escape probability approximation; these effects are unimportant for plausible [\oiii] velocity distributions).

Under these assumptions, the luminosity in the [\oiii] 88 $\mu$m line ($L_{10}$) and the 52 $\mu$m luminosity ($L_{21}$) may be written as:
\begin{equation}\label{eq:model}
    L_{ij}=\dfrac{R_i}{1+R_1+R_2}\left(\dfrac{n_\mathrm{O}}{n_\mathrm{H}}\right)_\odot\dfrac{Z}{Z_\odot}\dfrac{A_{ij}}{1+0.5\tau_{ij}}h\nu_{ij}\dfrac{Q_\mathrm{HI}}{\alpha_\mathrm{B,HII}n_e}\,.
\end{equation}
Here $R_i$ is the fractional abundance of OIII ions in the i-th energy state and $(n_\mathrm{O}/n_\mathrm{H})_\odot=10^{-3.31}$ is the solar oxygen to hydrogen abundance ratio. The quantity $A_{ij}$ is the Einstein-A coefficient, specifying the spontaneous decay rate from the i-th to the j-th energy level. The optical depth, $\tau_{ij}$, is treated self-consistently in the escape probability approximation but is unimportant in practice. Here $\nu_{ij}$ is the rest-frame frequency of the corresponding [\oiii] emission line, $\alpha_\mathrm{B,HII}$ is the case B recombination rate of hydrogen, and $n_e$ is the number density of free electrons. This formula then connects the luminosity in each line to the ISM parameters, $n_{\mathrm{H}}$ and $Z$. 

\begin{figure}
    \centering
    \includegraphics[width=0.4\textwidth]{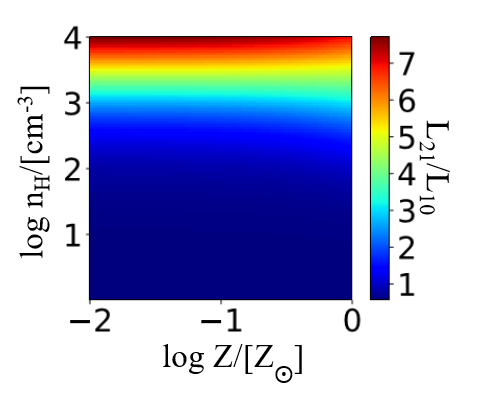}
    \caption{The luminosity ratio in the 52 $\mu$m to 88 $\mu$m lines for different gas densities and metallicities. The ionizing photon rate is fixed at $Q_\mathrm{HI}=10^{54}\ \mathrm{s^{-1}}$ (expected for an SFR of roughly $\sim 10 M_\odot/\mathrm{yr}$) and the gas temperature is set to $T=10^4$ K here. The line ratio provides a density indicator.}\label{fig:Rmap}
\end{figure}

 Yang2020 used this analytic model to constrain the ISM properties of nine ALMA targets with $L_{10}/\mathrm{SFR}$ observations.
 As discussed earlier, this left a degeneracy between gas density and metallicity (see also Figure~\ref{fig:posterior}). This can be broken by adding [\oiii] 52 $\mu$m measurements, owing to the different critical densities of the two lines. It is instructive to examine the asymptotic behavior of the line ratio, as discussed in Yang2020:
\begin{equation}\label{eq:ratio}
\dfrac{L_{21}}{L_{10}}\\
=\begin{cases}
\dfrac{k_{02}}{k_{01}+k_{02}}\dfrac{\nu_{21}}{\nu_{10}}\approx 0.55 & \text{if $n_e\rightarrow0$}; \\
\\
\dfrac{g_2}{g_1}\dfrac{A_{21}}{A_{10}}\dfrac{\nu_{21}}{\nu_{10}}=10.71 & \text{if $n_e\rightarrow\infty$},
\end{cases}
\end{equation}
where $g_2=5$ and $g_1=3$ are the degeneracies of the $^3\mathrm{P}_2$ (``2'') and $^3\mathrm{P}_1$ (``1'') levels. 
The $k$s are OIII collisional excitation rates. The low-density limit has a slight temperature dependence, but this is weak since the energy separation between these states is small relative to the HII region temperature. The number given in Eq~\ref{eq:ratio} assumes a gas temperature of $T=10^4$ K. To further illustrate, Figure~\ref{fig:Rmap} shows the line ratio across the $\log n_\mathrm{H}-\log Z$\footnote{In this work $\log$ denotes a base-10 logarithm.} parameter space assuming $Q_\mathrm{HI}=10^{54}$ s$^{-1}$ and $T=10^4$ K. The luminosity ratio transitions between the low and high density limits of Eq~\ref{eq:ratio} with a relatively sharp increase between $10^2 \lesssim n_{\mathrm{H}} \lesssim 10^3$ cm$^{-3}$, above which the 52 $\mu$m line is more luminous. The ratio is almost independent of metallicity, with only a weak dependence from the fact that the number density of free electrons depends slightly on $Z$. There is also a small effect that arises because the gas temperature depends on metallicity (see Yang2020 Eq 2); this is not captured in Figure~\ref{fig:Rmap}, which adopts a fixed temperature, but is included in our modeling. 

\par

Using the $n_\mathrm{H}-Z$ constraints in Yang2020, determined from the $L_{10}/\mathrm{SFR}$ measurements in the literature, we identify the currently allowed range in $L_{21}/L_{10}$ for each of the ten ALMA galaxies. The maximum and minimum values of $L_{21}/L_{10}$ allowed at 68\% confidence after spanning the $n_\mathrm{H}-Z$ parameter space are given in column (6) of Table~\ref{tb:data} for each target. In the alternate case, motivated earlier,  that the stellar metallicity ($Z_\star$) is 0.2 times the gas-phase metallicity ($Z$), the higher rate of ionizing photon production in this scenario increases the $L_{21}/L_{10}$ upper bounds by 5\%-28\%, depending on the target.

\section{Observing time}\label{sec:OT}

We now turn to compute the expected SNR for future ALMA 52 $\mu$m observations. We determine the ALMA sensitivity for a 10-hour measurement as well as the on-source observing time required to achieve a SNR$=6$ [\oiii] 52 $\mu$m detection at peak flux density for each target under different $L_{21}/L_{10}$ scenarios. Specifically, we consider the minimum and maximum $L_{21}$ allowed by the Yang2020 constraints at 68\% confidence, as well as an average between these two luminosity limits. 

In order to compute the ALMA sensitivity for each of the ten targets in Table~\ref{tb:data} we use the ALMA Sensitivity Calculator web interface.
We assume a Gaussian line profile for each emission line and that the full-width-at-half-maximum (FWHM) of the 52 $\mu$m line matches the best fit observational line width of its 88 $\mu$m counterpart. We adopt the 43 ALMA antenna configuration and frequency channels of width set by the FWHM/3, such that each spectral line is resolved by three channels. We furthermore assume that the angular size of the target is smaller than the beam of the observations, i.e., that the observations do not resolve the target. For simplicity we set the declination as zero for all of the sources.
The ALMA band containing the 52 $\mu$m line for each target is listed in column (5) of Table~\ref{tb:data}. The resulting 1-$\sigma$ noise on the integrated 52 $\mu$m luminosity for a 10-hour on-source measurement is given in column (7). Finally, the on-source integration times required for  6-$\sigma$ detections are shown for the maximum, mean, and minimum $L_{21}$ cases in column (8). As mentioned in \S \ref{sec:3}, decreasing $Z_\star$ at fixed gas phase metallicity increases $L_{21}/L_{10}$ and so shorter integration times are required for 52 $\mu$m detections in this case. We are therefore providing conservative detection time estimates.

Among the ten targets studied in this work, the 52 $\mu$m line emitted by MACS1149-JD1 is not observable by ALMA because its observed 52 $\mu$m frequency falls in the gap between the band 8 and band 9 windows. The other nine targets are in principle observable. The atmospheric opacity is, however, very large at the observed frequencies of the 52 $\mu$m line for A2744-YD4, MACS0416-Y1, and BDF-3299 and so it is not feasible to detect these objects in practice. If the galaxies in this data set tend to have high gas densities, $n_\mathrm{H} \gtrsim 10^2-10^3$ cm$^{-3}$, then the [\oiii] 52 $\mu$m lines from SXDF-NB1006-2, B14-65666, J0217-0208, and J1211-0118 can be detected within 10 hours, in agreement with the earlier work of Jones2020. As noted in Jones2020, modeling of [\oii] and [\sii] doublet emission lines in the spectra of LBGs at $z \sim 2-3$ indicates gas densities in the $n_\mathrm{H} \sim 200-300$ cm$^{-3}$ range \citep{2016ApJ...816...23S,2016ApJ...826..159S,Strom:2016nae}. If the ALMA galaxies are similar, or still more dense -- as one might expect for galaxies at higher redshifts -- then this bodes well for detecting their 52 $\mu$m emission. \footnote{The listed times are on-source observing times only, ignoring the calibration overheads. Furthermore, observations in ALMA band 9 and 10 require excellent weather conditions with good phase-stability. In practice, the effective band 9 and 10 observing time per local siderial time hour is limited to just a few hours per ALMA configuration. This is especially true for the compact configurations offered in Cycle 8 2021, the configurations most suitable for a detection experiment.} 

At low gas densities, the 52 $\mu$m emission line is significantly harder to detect and it may be possible to place only upper limits on the luminosity of this line. Nevertheless, even in the low density case where the 52 $\mu$m line is almost a factor of 2 less luminous than the 88 $\mu$m emission, J0217 is still theoretically detectable within 24 hours of on-source integration time. In the intermediate density case (where we take $L_{21}=L_{10}\times((L_{21}/L_{10})_\mathrm{max}+(L_{21}/L_{10})_\mathrm{min})/2$), we find that 6-$\sigma$ detections are possible in less than 10 hours for SXDF-NB1006-2 and B14-65666. In the high and intermediate density cases, SXDF-NB1006-2 requires the least time for a significant detection ($\sim 2-6$ hours). We therefore suggest SXDF-NB1006-2 as a promising first target for an ALMA 52 $\mu$m follow up measurement.\par

\section{52 $\mu$m measurement and ISM parameter constraints}\label{sec:5}

\begin{figure*}
    \centering
    \includegraphics[width=1\textwidth]{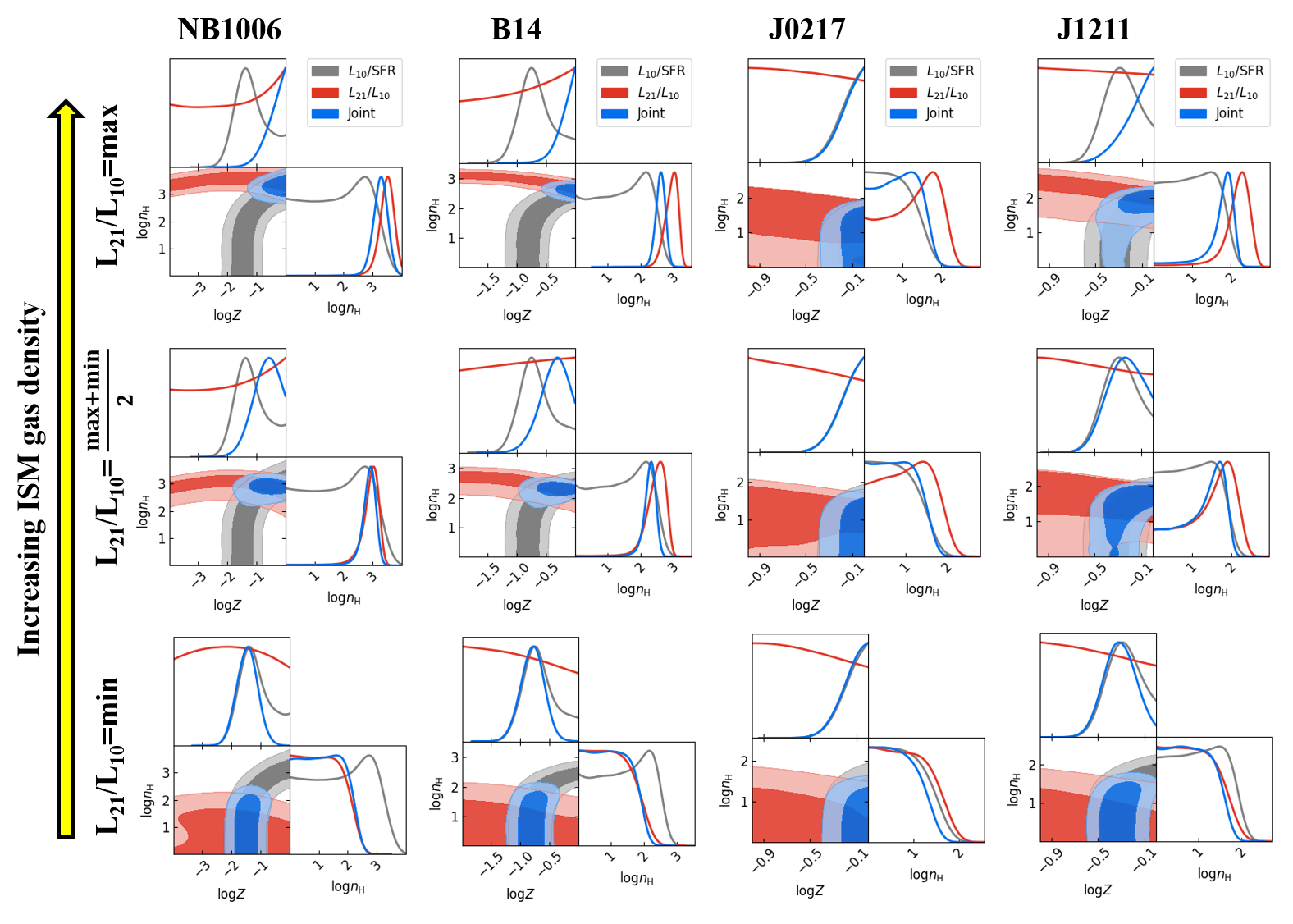}
    \caption{Forecasted improvements on gas density-metallicity parameter constraints from combining current $L_{10}/\mathrm{SFR}$ measurements with upcoming redshifted 52 $\mu$m observations in four example galaxies,  SXDF-NB1006-2, B14-65666, J0217-0208, and J1211-0118 (see Table~\ref{tb:data}). In each panel, the grey regions and lines show the constraints from the current $L_{10}/\mathrm{SFR}$ measurements, the red regions and lines show forecasts for the 52 $\mu$m line detections alone, while the blue regions and lines give the joint constraints that will be possible. The upper/middle/lower rows show results for the maximum/average/minimum $L_{21}$ scenarios (see text).  The shaded regions give 68\% and 95\% confidence intervals. 
    }\label{fig:posterior}
\end{figure*}

Figure~\ref{fig:posterior} shows forecasts for how the parameter constraints in the $\log n_\mathrm{H}-\log Z$ plane will improve after including $L_{21}/L_{10}$ measurements. Specifically, we select the four most promising targets from Table~\ref{tb:data}: SXDF-NB1006-2, B14-65666, J0217-0208, and J1211-0118 and consider 10 hour on-source integration times for the 52 $\mu$m followup observations (see column (7) of Table~\ref{tb:data} for the resulting noise estimates.) These forecasts are combined with the constraints obtained in Yang2020 from the $L_{10}/\mathrm{SFR}$ measurements of \cite{harikane2020large} (grey regions in the figure). In each case, we adopt a hard metallicity prior enforcing $Z \leq Z_\odot$. 

As in the previous section, we explore the constraints expected for three different assumptions regarding the line ratio $L_{21}/L_{10}$. We assume a Gaussian likelihood for the line ratio, the standard error propagation formula to compute the uncertainties on this ratio from the independent luminosity measurement errors, and MCMC calculations to forecast the expected parameter constraints.  
The top row of Figure~\ref{fig:posterior} shows the maximal case, the middle row the intermediate scenario, and the bottom row gives the minimum line ratio model calculations. The forecasted constraints from the line ratio alone are shown by the red regions/lines in Figure~\ref{fig:posterior}, while the combined constraints are given in blue. The variations from bottom to top correspond to increasing the gas density in the ISM, as discussed earlier and indicated by the yellow arrow in the figure.

As noted previously, the $L_{10}/\mathrm{SFR}$ measurements alone leave a strong degeneracy between gas density and metallicity, giving rise to the ``L''-shaped grey regions in the figure. The 52 $\mu$m line ratio measurements will add nearly horizontal ellipses in the $\log n_\mathrm{H}-\log Z$ plane through their sensitivity to the gas density. The three cases show the impact of shifting the fiducial gas density from a higher value to a lower value as one moves from the top row to the bottom one. Note that in some examples these constraint forecasts come only from upper limits on the 52 $\mu$m line emission: for the minimum $L_{21}/L_{10}$ case we do not expect 6-$\sigma$ line detections towards any of the four targets in less than ten hours. Further, for the intermediate $L_{21}/L_{10}$ model, J0217-0208 and J1211-0118 still fall below the SNR=$6$ threshold. Those examples illustrate that even an upper limit will help to constrain the parameter space here.

In most cases, we forecast that the gas density posteriors will tighten significantly after including 52 $\mu$m follow-up measurements. In many of the examples shown this should also help to tighten the metallicity constraint by breaking the degeneracy between density and metallicity. For instance, in the maximal line ratio case it will be possible to decisively show that SXDF-NB1006-2 has a high metallicity $Z \geq 0.092Z_\odot$ at 95\% confidence, while based on $L_{10}/\mathrm{SFR}$ alone this galaxy's metallicity may be as low as $Z=0.011Z_\odot$, again at 95\% significance. The improvements forecast are weakest for the case of J0217-0208. This galaxy has a very large $L_{10}/\mathrm{SFR}$ and so high densities are already excluded for this system (at least under our $Z \leq Z_\odot$ prior). Nevertheless, we expect the gas density determination to improve for this galaxy although the metallicity constraint will not tighten. Overall, the prospects for obtaining more stringent ISM parameter constraints from 52 $\mu$m observations towards several of the galaxies in the current sample appear promising. In addition, we expect that further 88 $\mu$m detections will extend the list of promising targets here in the near future.\footnote{For example, the ALMA Large Program REBELS (2019.1.01634.L) aims to discover the most luminous [\cii] and [\oiii] galaxies in the EoR.}  

\par

\section{52 micron measurements and the mass-metallicity relation}\label{sec:MZ}

The previous section shows that 52 $\mu$m measurements can help sharpen constraints on the gas density and metallicity. Here we turn to further explore the potential scientific impact of these measurements, by quantifying the improvements that will be possible in our understanding of the mass-metallicity relationship during the EoR. This will extend the work of Jones2020, which started to constrain this important relationship from the current [\oiii] 88$\mu$m measurements. As mentioned in the Introduction, the improved gas density limits may also help in understanding the escape fraction of ionizing photons. Exploring this, however, will likely require comparison with detailed radiation-hydrodynamical simulations capable of modeling the escape fraction and its dependence on galaxy properties, and so we defer this to possible future work. 

\begin{figure*}
    \centering
    \includegraphics[width=1.0\textwidth]{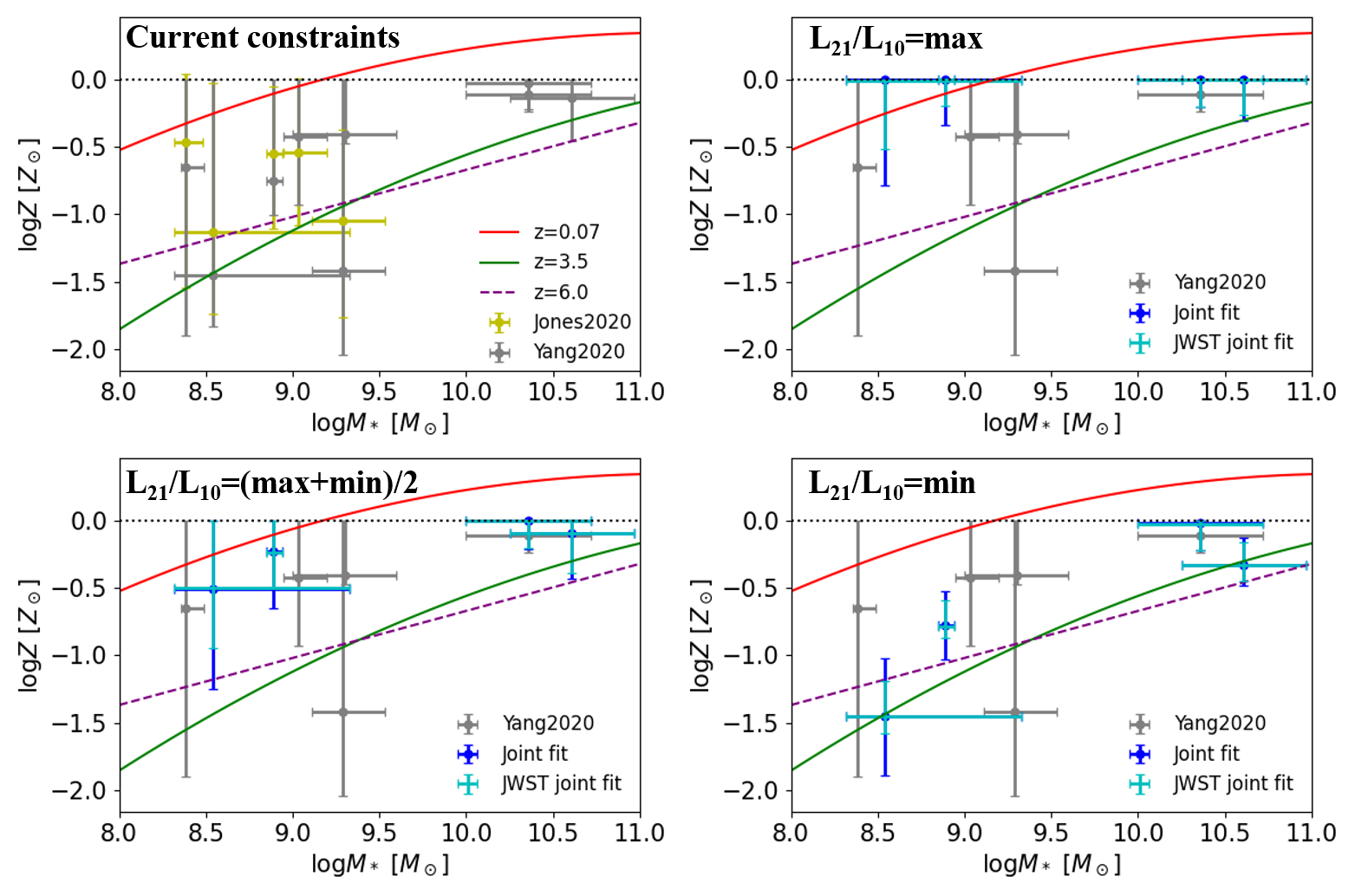}
    \caption{
    Constraints on the mass-metallicity relationship from Jones2020 and Yang2020, compared with results from simulations and observations in the literature. In each panel, the mass-metallicity evolution from $z\sim 0$ to $z \sim 4$ determined empirically by the \textsc{AMAZE} project (Maiolino2008) is shown (two thin curves at $z=0.07$ and $z=3.5$.). The results from the \textsc{FIRE} simulation suite (Ma2016) at $z=6$ are given by the purple dashed line.  In each panel, the black dotted line at solar metallicity indicates the metallicity prior adopted in Yang2020 and this work. {\em Top left}: The metallicity constraints from Jones2020 and Yang2020 are shown as yellow and grey crosses, respectively. The other three panels illustrate
    how the error bars in the mass-metallicity plane may tighten
    with the addition of 52 $\mu$m measurements. The top right, bottom left, and bottom right panel correspond to the maximum/average/minimum $L_{21}$ scenarios (see text). 
    In each case, we show results for NB1006, B14, J0217, and J1211, identified as promising targets in the previous section, in blue crosses. The cyan points further show improvements that may be possible with upcoming JWST SFR measurements, provided the resulting $L_{10}/\mathrm{SFR}$ errors become dominated by the uncertainties on $L_{10}$ alone.
    Metallicity constraints for JD1, YD4, Y1, BDF, J0235 from Yang2020 are shown in grey crosses (these galaxies are less promising for follow-up 52 $\mu$m measurements.) 
    All of the error bar in this figure are $1-\sigma$ uncertainties.
    }\label{fig:MZ}
\end{figure*}\par

The top left panel of Figure~\ref{fig:MZ} compares empirical fits to the mass-metallicity relationship at lower redshifts and a fitting formula from a current simulation in the literature, along with the 88 $\mu$m based metallicity determinations from each of Jones2020 and Yang2020 at $6 < z \lesssim 9$. More specifically, the best-fit empirical $Z(M_*)$  relationships at $z \sim 0$ and $z=3.5$ from \cite{2008A&A...488..463M} (hereafter Maiolino2008) are shown in solid curves. These are compared to results from the FIRE simulations (dashed line), calibrated from simulated galaxies at $z=6$ \citep{2016MNRAS.456.2140M} (hereafter Ma2016). Note that the Ma2016 $Z(M_*)$ fitting formula predicts a drop of only 0.05 dex between $z=6$ and $z=9$; this estimate requires extrapolating their formula beyond the redshift range in which it was calibrated, at $z \leq 6$, out to the highest redshift ($z \sim 9$) in the current ALMA sample. Hence the evolution across the redshifts of this data set is expected to be negligibly small compared to present and upcoming measurement errors. 

The current $6 < z \lesssim 9$ metallicity constraints from Jones2020 and Yang2020, along with the stellar mass measurements in Table~\ref{tb:data}, are given by the yellow and grey points, respectively, each with $1-\sigma$ error bars. As discussed further in Yang2020, these two studies agree despite different methodologies, although the error bars in the Yang2020 work are larger, given the more agnostic prior on density adopted in that work. Note also that the sample considered in Yang2020 includes additional galaxies beyond those in Jones2020: BDF-3299, J0217-0208, J0235-0532, and J1211-0118.

In order to quantify the constraints on the mass-metallicity relation from Jones2020 and Yang2020, we calculate the reduced $\chi^2$ difference between the measurements and model $Z(M_*)$ 
relationships. In doing so, it is important to account for the intrinsic galaxy-to-galaxy scatter in this relationship, which is not well characterized in the current \text{AMAZE} observations of Maiolino2008 nor in the FIRE simulations of Ma2016. We therefore adopt an intrinsic scatter of $\sigma^{\mathrm{int}}=0.2$ dex based on results from another simulation study, using the IllustrisTNG simulations \citep{2019MNRAS.484.5587T}. Specifically, we denote the metallicity measurements towards the $i$th galaxy by 
$Z_i$ and the corresponding metallicity error as $\sigma^\mathrm{\log Z}_i$. The fitted Maiolino2008 and Ma2016 $Z(M_*)$ relations are labeled as $\log Z^\mathrm{model}$. The value of $\chi^2$ per degree of freedom, $\chi^2_\nu$, is then calculated as:
\begin{equation}\label{eq:chi2}
    \chi^2_\nu=\sum\limits_{i=1}^{N=9}\left(\dfrac{(\log Z_i-\log Z^\mathrm{model}_i)^2}{(\sigma^\mathrm{\log Z}_i)^2+(\sigma^\mathrm{int}_i)^2+f'(\log M_*)^2(\sigma_i^{\log M_*})^2}\right)\dfrac{1}{N}\,,
\end{equation}
where the sum runs over the nine ALMA [\oiii] 88 $\mu$m detections.
In order to account for the uncertainties in the stellar mass estimates, we include a term in the variance that depends on the local derivative of the $Z(M_*)$ relationship with respect to the logarithm of stellar mass \citep{doi:10.1080/00401706.1967.10490460}, denoted above as $f'(\log M_*)$. This is calculated assuming the Ma2016 $Z(M_*)$ relationship as $f'(\log M_*)=0.35$. 

The comparison with current constraints from Jones2020 and Yang2020 indicates that none of the mass-metallicity relationships in the literature are strongly favored or disfavored at the moment. Quantitatively, the $\chi^2_\nu$ values for the example relationships in Figure~\ref{fig:MZ} are given in the first two columns of Table~\ref{tb:chi2}. The maximum $\chi^2_\nu$ values for Jones2020 and Yang2020 are 0.88 and 1.23, respectively, corresponding to minimum p-values of 0.49 and 0.27, and so the current measurement errors are too large to distinguish models. 

However, 52 $\mu$m measurements may help here, although the constraint obtained will depend on the gas densities in the target galaxies. Here we only consider the additional metallicity constraints contributed by 52 $\mu$m measurements of the most promising targets: NB1006, B14, J0217, and J1211, as identified in \S\ref{sec:OT}. The top right panel of Figure~\ref{fig:MZ}, for instance, illustrates that if the gas density in the four promising ALMA targets is large (so that $L_{21}/L_{10}$ is near the maximal case identified in the previous section), the high metallicity implied in these galaxies would disfavor the simulated mass-metallicity relation in Ma2016 and the higher redshift empirical fit from Maiolino2008. That is, in this case, the data may prefer less redshift evolution in the normalization of the mass-metallicity relationship than in the \textsc{AMAZE} sample and in the FIRE simulations. 
On the other hand, in the minimal $L_{21}/L_{10}$ scenario (bottom right) the lower redshift mass metallicity relationships of Maiolino2008 would be modestly disfavored. In the intermediate case (lower left) the results may not strongly discriminate between any of the models shown. 

\begin{table*}
\centering
\begin{tabular}{lcccccccc}
\hline
&&&& \\[-1em]
$Z(M_*)$&Jones2020&Yang2020&max $L_{21}/L_{10}$&mean $L_{21}/L_{10}$&min $L_{21}/L_{10}$&max $L_{21}/L_{10}$&mean $L_{21}/L_{10}$&min $L_{21}/L_{10}$\\
&&&& \\[-1em]
Relations&&&ALMA&ALMA&ALMA&JWST&JWST&JWST\\
&&&& \\[-1em]
\hline
&&&&\\[-1em]
$z\sim0$& 0.88,0.49&1.23,0.27&0.97,0.46&1.08,0.37&2.27,1.5E-2&0.99,0.45&1.08,0.37&2.86,2.2E-3  \\
&&&&\\[-1em]
$z=3.5$&0.55,0.74&0.81,0.61&2.81,2.6E-3&1.99,3.5E-2&1.50,0.14&4.28,1.39E-5&2.70,3.8E-3&1.71,8.1E-2   \\
&&&&\\[-1em]
$z=6.0$&0.38,0.86&1.48,0.15&2.85,2.3E-3&2.12,2.4E-2&1.71,8.1E-2&3.97,4.4E-5&2.61,5.2E-3&1.83,5.8E-2\\
&&&&\\[-1em]
\hline
\end{tabular}
\caption{Summary of the reduced $\chi^2$ test results. The columns specify the mass-metallicity constraints/forecasts discussed in this work, while the rows show varying $Z(M_*)$ models. The $z \sim 0$ and $z=3.5$ cases are empirical relations from Miaolino2008, while the $z=6.0$ model is from the FIRE simulations of Ma2016. The second and third columns correspond to the current constraints, while the remaining ones give forecasts in various scenarios (see text). 
The first entry in each cell is the (reduced) $\chi^2_\nu$ value calculated following Eq~\ref{eq:chi2}. The second entry specifies the corresponding p-value.}\label{tb:chi2}
\end{table*}

In order to quantify these qualitative trends and our ability to constrain the mass-metallicity relationship with the upcoming 52 $\mu$m measurements, we calculate $\chi^2_\nu$ between mock metallicity measurements in each scenario and the different empirical/simulation models. In computing $\chi^2_\nu$ with Eq~\ref{eq:chi2} we use here the joint metallicity constraints from the current $L_{10}/\mathrm{SFR}$ and mock $L_{21}/L_{10}$ measurements for NB1006, B14, J0217, and J1211, while we use the constraints from Yang2020 for the other fixe ALMA [\oiii] 88 $\mu$m targets. The $\chi^2_\nu$ test results are summarized in the fourth through sixth column in Table~\ref{tb:chi2}.
Most notably, in the `$L_{21}/L_{10}=\mathrm{max}$' scenario, the $\chi^2_\nu$ forecasts for the Maiolino2008 $z=3.5$ model and Ma2016 $z=6.0$ model are 2.81 and 2.85 respectively, corresponding to p-values on the order of $10^{-3}$. That is, these cases would be strongly -- if not decisively -- disfavored. 

Furthermore, as discussed in Jones2020, future JWST measurements of recombination lines such as H$\alpha$ and H$\beta$ should allow improved SFR determinations and dust extinction corrections. To gauge the improvements that may be possible, we suppose that the $L_{10}/\mathrm{SFR}$ measurement errors become dominated by $\sigma_{L_{10}}$ alone. In this case, the forecasted $\chi^2_\nu$ values increase to 4.28 and 3.97 for the maximal mock [\oiii] 52 $\mu$m signal scenario, corresponding to p-values of  $1.4\times10^{-5}$ and $4.4\times10^{-5}$. That is, in this case, the high redshift Maiolino2008 and Ma2016 models can be decisively excluded. The prospects in other scenarios are summarized in columns seven to nine of Table~\ref{tb:chi2}. In addition, JWST will be able to measure rest-frame optical emission lines from [\oiii] and [\oii]. The [\oiii] optical lines are sensitive to the temperature of the emitting gas (e.g. Jones2020 and Yang2020) and so are likely less robust metallicity indicators on their own. However, in combination with the 88 $\mu$m and 52 $\mu$m transitions, the rest-frame optical [\oiii] and [\oii] lines can help in empirically determining gas temperatures and ionization states.  See Jones2020 for further discussion. 
Hence the future combination of 52 $\mu$m and JWST SFR measurements should help in determining the mass-metallicity relationship during the EoR. It will also be important, however, to obtain a significantly larger sample of targets. In this context, galaxies with relatively low stellar masses, such as NB1006-2, seem especially interesting since measurements at the low mass end will be valuable in determining the {\em shape} of the $Z(M_*)$ relationship at high redshift. 

In the nearer term, these results suggest that the upcoming measurements will start to place interesting constraints on the evolution in the {\em normalization} of the $Z(M_*)$ relationship. Since the current data are consistent with a range of possibilities, from surprisingly little evolution -- in comparison with even the $z \sim 0$ normalization from the \textsc{AMAZE} sample  -- to strong evolution (e.g. the current measurements are consistent with models below the Ma2016 $z \sim 6$ normalization), it is perhaps premature to speculate on the precise implications in different scenarios. Instead, we only remark that the $Z(M_\star)$ normalization depends on the combined impact of evolving gas fractions, the efficiency of metal retention, and metal yields; it therefore provides a rich test of models of galaxy formation and feedback in a redshift regime well beyond that in which the models have been calibrated (e.g. Ma2016, \citealt{2019MNRAS.484.5587T}).

\section{Conclusion}\label{sec:Conclusion}

We have forecast the prospects for detecting the [\oiii] 52 $\mu$m emission line from the current sample of ten ALMA [\oiii] 88$\mu$m measurements towards galaxies at $6 < z \lesssim 9$. Adding 52 $\mu$m detections or upper limits will break the degeneracy between gas density and metallicity that remain from 88 $\mu$m and SFR measurements alone. Using the Yang2020 model we forecast that the [\oiii] 52 $\mu$m lines from SXDF-NB1006-2, B14-65666, J0217-0208, and J1211-0118 can be detected within 10 hours of on-source observing time, provided these galaxies have gas densities larger than $n_{\mathrm{H}} \gtrsim 10^2-10^3$ cm$^{-3}$.

We forecast the parameter space improvements that will be possible for the four most promising targets, SXDF-NB1006-2, B14-65666, J0217-0208, and J1211-0118, with a ten hour ALMA 52 $\mu$m measurement. We find that the gas density constraint should tighten by 1-3 dex for these sources, while the metallicity posterior will narrow by as much as 1 dex for SXDF-NB1006-2, B14-65666, and J1211-0118. We identify SXDF-NB1006-2 as the most favorable target for first followup observations. The 52 $\mu$m measurements will enable interesting constraints on the mass metallicity relationship during the EoR, especially if these target galaxies have high gas densities. Further improvements are expected here following better SFR determinations from the JWST.

The 52 $\mu$m measurements should also help us understand the gas density of some of the first galaxies. These observations will help determine the internal structure of high redshift galaxies and photon escape fraction in the sources that reionized the universe. 

\section{Acknowledgement}

AL acknowledges support through NASA ATP grant 80NSSC20K0497. We thank  the referee for helpful suggestions, especially related to Section 6 of the manuscript.

\section*{Data availability}
The data used to support the findings of this study are available from the corresponding author upon request.



\bibliographystyle{mnras}
\bibliography{OIII}




\label{lastpage}
\end{document}